\documentclass[fleqn,usenatbib]{mnras}

\usepackage[T1]{fontenc}
\usepackage{ae,aecompl}

\usepackage{graphicx}	
\usepackage{amsmath}	
\usepackage{amssymb}	
\usepackage{ragged2e}
\usepackage{booktabs}   
\usepackage{multirow}   

\newcommand{\epse}{\varepsilon_{e}}
\newcommand{\epsB}{\varepsilon_{B}}

\newcommand{\thetaobs}{\theta_v}

\title[GW170817 in X-rays]{A thousand days after the merger: continued X-ray emission from GW170817}  

\author[Troja et al.]{
 E.~Troja$^{1,2}$ \thanks{E-mail: eleonora.troja@nasa.gov},
H.~van~Eerten$^{3}$,
B.~Zhang$^{4}$, 
G.~Ryan$^{1,5}$, 
L.~Piro$^{6}$,
R.~Ricci$^{7,8}$,  
\newauthor
B. O'Connor$^{1,2,9,10}$, 
M.~H.~Wieringa$^{11}$, 
S.~B.~Cenko$^{2,5}$, 
T.~Sakamoto$^{12}$
\\
$^{1}$ Department of Astronomy, University of Maryland, College Park, MD 20742-4111, USA \\
$^{2}$ Astrophysics Science Division, NASA Goddard Space Flight Center, 8800 Greenbelt Rd, Greenbelt, MD 20771, USA \\
$^{3}$ Department of Physics, University of Bath, Claverton Down, Bath, BA2 7AY, UK \\
$^{4}$ Department of Physics and Astronomy, University of Nevada, 89154, Las Vegas, NV, USA \\
$^{5}$ Joint Space-Science Institute, University of Maryland, College Park, Maryland 20742, USA \\
$^{6}$ INAF, Istituto di Astrofisica e Planetologia Spaziali, via Fosso del Cavaliere 100, 00133 Rome, Italy \\
 $^{7}$ INAF-Istituto di Radioastronomia, Via Gobetti 101, I-40129, Bologna, Italy \\
 $^{8}$ Istituto Nazionale di Ricerca Metrologica (INRiM) - Strada delle Cacce 91 - Torino, Italy\\
 $^{9}$ Department of Physics, The George Washington University, 725 21st Street NW, Washington, DC 20052, USA\\
 $^{10}$ Astronomy, Physics, and Statistics Institute of Sciences (APSIS), The George Washington University, Washington, DC 20052, USA  \\
 $^{11}$ CSIRO Astronomy and Space Science, PO Box 76, Epping, New South Wales 1710, Australia \\
$^{12}$ Department of Physics and Mathematics, Aoyama Gakuin University, 5-10-1 Fuchinobe, Chuo-ku, Sagamihara-shi Kanagawa 252-5258, Japan
}

\date{Accepted XXX. Received YYY; in original form 2020 Jun 1}

\pubyear{2020}
\begin{document}
\maketitle

\begin{abstract}
Recent observations with the {\it Chandra} X-ray telescope continue to detect X-ray emission from the transient GW170817. In a total exposure of 96.6 ks, performed between March 9 and March 16 2020 (935~d to 942~d after the merger), a total of 8 photons are measured at the source position, corresponding to a significance of  $\approx$5$\sigma$. Radio monitoring with the 
Australian Telescope Compact Array (ATCA)
shows instead that the source has faded below our detection threshold ($<$33$\mu$Jy, 3$\sigma$). 
By assuming a constant spectral index of $\beta$=0.585, we derive an unabsorbed X-ray flux of $\approx$1.4$\times$10$^{-15}$\,erg\,cm$^{-2}$\,s$^{-1}$, higher than earlier predictions, yet still consistent with a simple structured jet model. We discuss possible scenarios that could account for prolonged emission in X-rays. 
The current dataset appears consistent both with energy injection by a long-lived central engine and with the onset of a kilonova afterglow, arising from the interaction of the sub-relativistic merger ejecta with the surrounding medium. Long-term monitoring of this source will be essential to test these different models. 

\end{abstract}



\section{Introduction}

On August 17th 2017, advanced LIGO and Virgo observed the first gravitational wave signal from a binary neutron star (NS) merger \citep{gw170817}. This event, named GW170817, was followed by a
short duration gamma-ray burst, GRB170817A,  and, 9 days later, by a non-thermal afterglow emission, visible across the electromagnetic spectrum \citep{grb170817, Troja17, Hallinan17}. 
After an initial rising phase, $F \propto t^{0.8}$ \citep{Troja18,Mooley18,Lyman18,Margutti18, Ruan18}, the afterglow peaked at $\approx$160 d after the merger and then started a rapid decay phase, $F \propto t^{-2.2}$ \citep{Mooley18b,Lamb19,Troja19}. This behavior is markedly different from the garden-variety GRB afterglows, observed to fade within a few minutes since the burst. 

The low-luminosity of the gamma-ray emission and the atypical temporal evolution of the afterglow component are widely interpreted as manifestation of a highly-relativistic structured jet seen at an angle of $\approx$20-30 deg from its axis \citep{Troja17,grb170817,Lazzati18,Lyman18,Troja18,Mooley18b,Margutti18,Lamb19,Ryan19,Troja19,Hajela19}. 
In this model, the energy and Lorentz factor of the relativistic ejecta vary with the angle from the jet's axis \citep[e.g.][]{ZM02}.
The initial rising slope and the peak time strongly depend on the observer's viewing angle and the jet's angular profile \citep{Ryan19}. However, the post-peak behavior is dominated by the emission from the jet's core and should resemble the post jet-break evolution of a standard GRB afterglow.
Even in this case, the post-break evolution can exhibit a rich behaviour, and is sensitive to the nature of the spreading dynamics of the decelerating relativistic plasma and to gradients in the circumburst ambient gas mass distribution. 
At sufficiently late times, emission from the jet as it has decelerated to non-relativistic flow velocities will begin to dominate the total observed flux, leading to a change in slope relative to the relativistic limit \citep{Frail00}. If a counter jet was launched, this too will at some point become visible \citep{vanEerten10}. However, very few GRBs are close enough to remain continuously visible for years and, for this reason, the jet's late-time evolution is rarely probed by observations at wavelengths other than radio \citep[e.g.][]{DePasquale16,Kouveliotou04}. 

Changes in the light curve evolution can also be the product of a genuinely new feature of the outflow not previously detected. Of particular interest to the case of neutron star mergers are scenarios that relate directly to the nature of the remnant (such as prolonged energy injection from a long-lived central engine, \citealt{Piro19}) and to the  sub-relativistic merger ejecta, producing a low-luminosity late-peaking afterglow \citep{np11, Hotokezaka18,Kathi18}.
In the case of GW170817, evidence suggests that a substantial amount ($\gtrsim$0.01\,$M_{\odot}$) of fast ($\gtrsim$0.1\,$c$) ejecta comes from the luminous kilonova emission AT2017gfo \citep{Arcavi2017, Evans2017, Drout2017, Kasen17, Kasliwal2017, Nicholl2017, Pian2017, Shappee2017, Smartt2017,  Tanvir2017, Troja17}. 
As these ejecta continue to expand they will drive a blastwave in the local medium, begin decelerating as more mass is swept up, and emit synchrotron radiation from the blast wave’s forward shock. This emission, which we refer to as kilonova afterglow, peaks years after the initial burst and,
at the distance of GW170817, may be bright enough to be detected with current instruments.

In order to explore the late-time behavior of the relativistic jet and constrain alternative components of emission, the location of GW170817 is periodically monitored at radio and X-ray energies. 
In this work, we present the results of the long-term monitoring campaign
with the {\it Chandra} X-ray observatory and the Australian Telescope
Compact Array (ATCA),  and discuss the possible origins of the observed long-lived X-ray emission. 
Throughout this paper, we adopt a distance of 40~Mpc and a standard $\Lambda$CDM cosmology \citep{Planck18}. Unless otherwise stated, the quoted errors are at the 68\% confidence level, and upper limits are at the 3\,$\sigma$ confidence level.

\begin{figure}
\includegraphics[width=0.99\columnwidth]{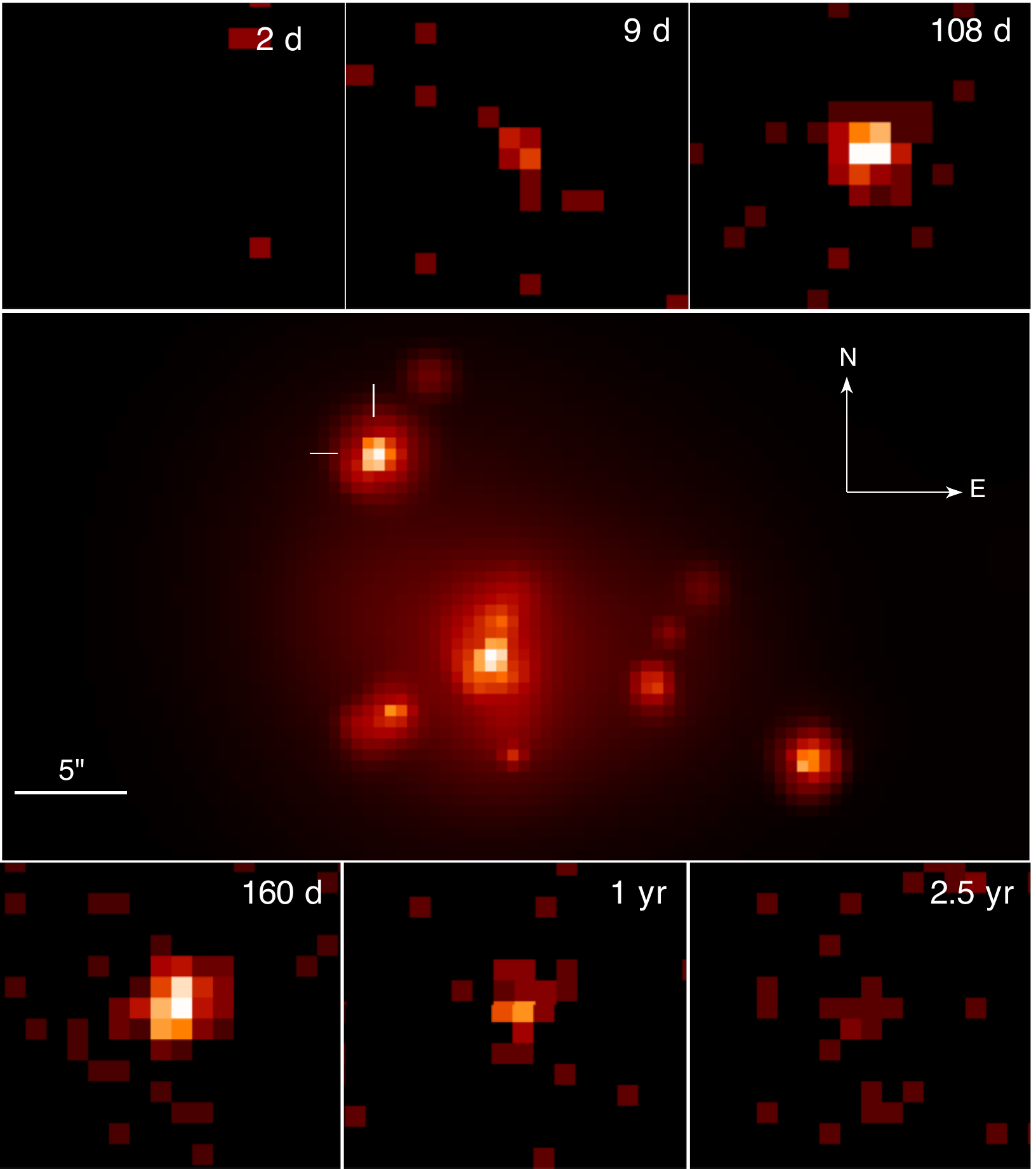}
\vspace{-0.2cm}
\caption{X-ray image of GW170817, as observed by {\it Chandra}.
The central pane shows the stacked image of the field, 
with total exposure of 783~ks. The image was adaptively smoothed
with a Gaussian kernel. 
The position of GW170817 is marked. In addition, several X-ray point sources as well as extended diffuse X-ray emission 
are visible. 
The image stamps are centered on the location
of GW170817, showing the main phases of its evolution. 
}
\label{fig:image}
\vspace{0.3cm}
\includegraphics[width=0.99\columnwidth]{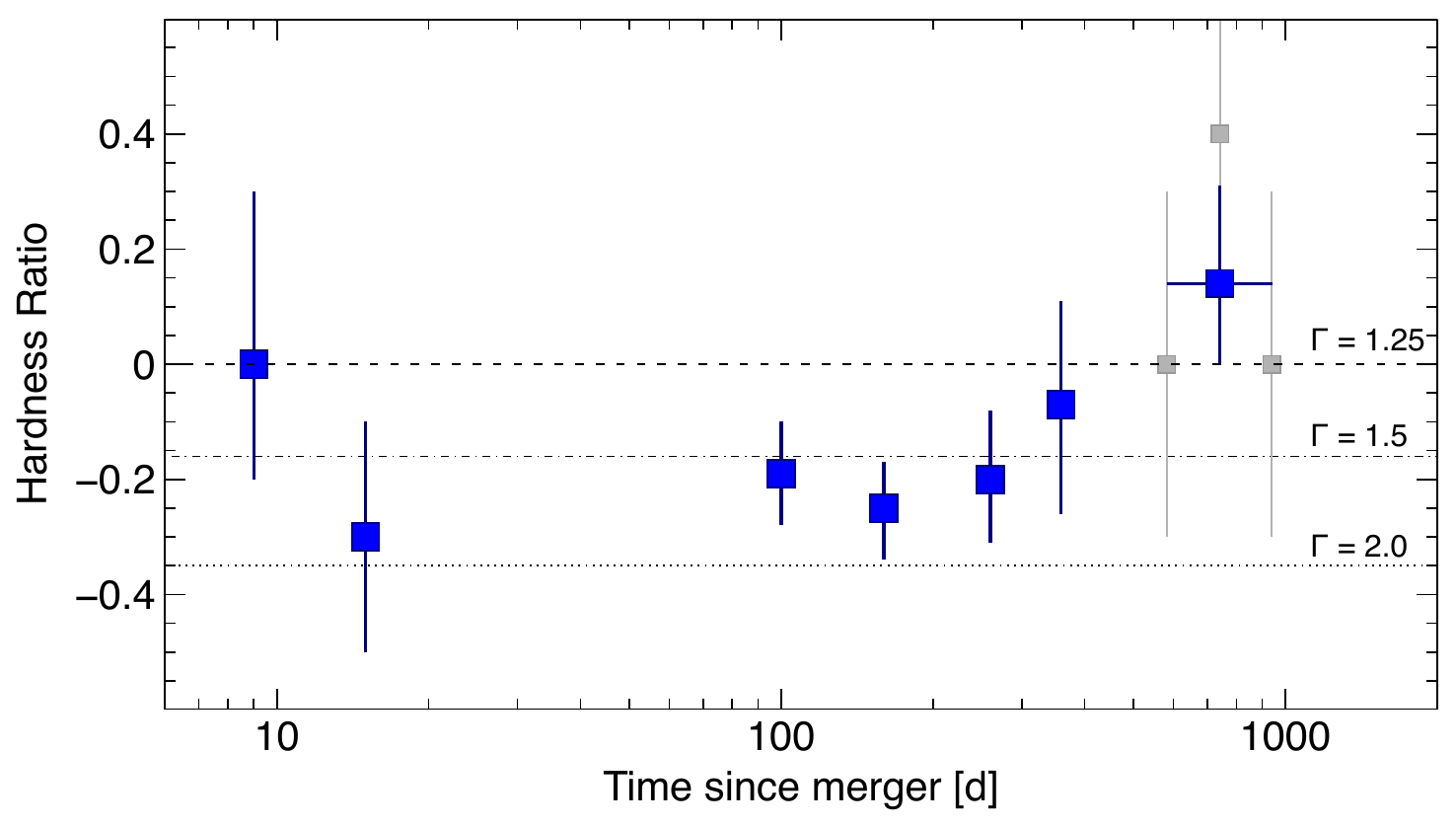}
\vspace{-0.3cm}
\caption{Hardness Ratio light curve for the X-ray afterglow of GW170817. 
We adopted the definition $HR = (H-S)/(H+S)$, where $H$ and $S$ are the net source counts in the hard (2.0-7.0 keV) and soft (0.5-2.0 keV) energy bands, respectively. Error bars represent 1\,$\sigma$ uncertainties. The last three epochs (gray symbols) were binned into a single point in order to improve the signal-to-noise ratio.
Horizontal lines show the values expected for an absorbed power-law with 
photon index $\Gamma$=2.0 (dotted line), 1.5 (dot-dashed line), and 1.25 (dashed line). 
}
 \label{fig:hr}
\end{figure}
\vspace{-0.2cm}

\begin{table*}[t]
        \centering
        \caption{Late-time X-ray observations of GW170817}
        \label{tab:obs}
        \begin{tabular}{lcccccc}
        \hline
        & T-T$_0$  & Exposure & Count Rate & Unabsorbed Flux &  Flux Density & Significance  \\
            &          &          & (0.5-7.0 keV) & (0.3-10 keV) &  5 keV & \\
        &  [d] & [ks] & [10$^{-4}$ cts s$^{-1}$] & [10$^{-15}$ erg cm$^{-2}$ s$^{-1}$] &  [10$^{-5}$ $\mu$Jy] & [$\sigma$]\\
         \hline
        Epoch 1 
        &  582 &  98.8 & 1.5 $\pm$ 0.4 & 2.6  $\pm$ 0.7 & 9 $\pm$ 2 & 7.7 \\
        Epoch 2 &  742 & 98.9  & 1.1$^{+0.4}_{-0.3}$ & 1.7$^{+0.7}_{-0.5}$ & 5.8 $\pm$ 1.7 & 6.1 \\
        Epoch 3\footnotemark
        &  939 & 96.6  & 0.8 $\pm$ 0.3   & 1.4 $\pm$ 0.5 &   4.7 $\pm$ 1.8  & 5.2  \\
      \hline
    \end{tabular}
\end{table*}

\section{Observations}
 \subsection{X-rays}\label{sec:Xrays}
We presented the analysis of the first year of observations in \citet{Troja19}. Since then, the target GW170817 is being monitored by the {\it Chandra} X-ray Telescope 
with a cadence of approximately six months under Guest Observer programs 20500691 (PI: Troja) and 20500299 (PI: Margutti). 
These three additional epochs (Table~\ref{tab:obs}) track the afterglow evolution from 1.6 to 2.6 years after the merger. The temporal evolution of the X-ray counterpart is shown in Fig.~\ref{fig:image}. 

Each epoch was split into multiple observations. Each observation was reduced in a standard fashion using the CIAO v4.12 and the latest calibration files (CALDB 4.9.1). 
In order to correct for small positional errors between different observations, we used the tool \textit{reproject\_aspect} to determine a new aspect solution based on common bright point sources. Each observation was reprocessed using the updated astrometric information. 
Data were filtered with the task \textit{deflare} to remove background flares by applying a sigma clipping threshold of 3. Observations carried out at a similar epoch were merged into a single image using the task \textit{flux\_obs}. The resulting total exposures are 98.8 ks (Epoch 1), 98.9 ks (Epoch 2), and 96.6 ks (Epoch 3). 

Aperture photometry was performed in the broad 0.5-7.0 keV energy band.
Source counts were extracted from the merged images using a circular aperture containing 92\% of the encircled energy fraction, whereas the background contribution was estimated from nearby source-free regions.
X-ray emission from the position of GW170817 is visible at all epochs. 
We estimated the detection significance following the Bayesian method of \citet{kbn}, and report in Table~\ref{tab:obs} the equivalent value for a normal probability distribution. 

Due to the low number of counts, the source spectral properties can not be adequately constrained. In order to check for possible spectral evolution we computed the hardness ratio (HR; \citealt{behr}), defined as the ratio $(H-S)/(H+S)$, where $H$ and $S$ are the net source counts in the hard (2.0-7.0 keV) and soft (0.5-2.0 keV) energy bands, respectively. The HR light curve (Fig.~\ref{fig:hr}) shows a possible hardening of the spectrum at late times 
($t\gtrsim$1.5~yr),
although with low significance. 

X-ray fluxes were calculated assuming an absorbed power-law spectrum with column density fixed to the Galactic value 1.1$\times$10$^{21}$\,cm$^{-2}$ \citep{Willingale13} and a photon  index $\Gamma$= $\beta$ +1 = 1.585, where $\beta$ is the spectral index derived from broadband afterglow modeling \citep{Troja19}. 
A harder spectrum would increase our flux estimate by $\approx$13\% (for $\Gamma$=1.25), still within the statistical uncertainties of the measurement. 
Our values are lower, yet consistent within the large uncertainties, than those reported in \citet{Hajela19}. Our conversion into fluxes is based on the broadband (from radio to X-rays) spectral shape and does not change over time, whereas \citet{Hajela19} derives variable conversion factors based on single-epoch X-ray observations. The latter approach is subject to greater uncertainty, and does not take into account the full spectral information available from the multi-wavelength dataset. 

\footnotetext
{An independent analysis of this data set reports
a similar count-rate and a 50\% higher X-ray flux \citep{Hajela20}. 
We can reproduce this result only by assuming 
a hard spectrum with $\Gamma$=0.57, 
drastically different from the 
spectral properties of the GW afterglow.}

\subsection{Radio}\label{sec:radio}

We re-observed the position of GW170817 with ATCA
(program C3240; PI:Piro) on May 3rd, 2020 (990 d since the merger) for 11 hours. The array configuration was 6A, the centre observing frequency was 2.1~GHz and the observing bandwidth was 2~GHz. The usual primary calibrator 1934-638 was not observed, instead the band-pass calibrator 0823-500 was used to bootstrap the absolute flux density scale assuming a flux density of 6.38~Jy and a spectral slope of -0.215. The source 1245-197 was used as the phase calibrator. The data set was calibrated and imaged in \texttt{Miriad} using standard procedures. The array configuration resulted in a E-W angular resolution of 6.5 arcsec, sufficient to separate the target from its host galaxy NGC~4493.  

No detection was found at the position of GW170817 in the natural-weighted restored image. A 3\,$\sigma$ upper limit of 33 $\mu$Jy was estimated from rms noise statistics in a region of the restored image away from bright radio sources. 
This measurement constrains the broadband spectral index
to $\beta$\,$<$0.68. 

\section{Model Fitting Methods}

Throughout this paper we continue our practice from \cite{Troja18, Troja19, Piro19, Ryan19} of performing Bayesian fits using the model and {\tt afterglowpy} software\footnote{\url{https://github.com/geoffryan/afterglowpy}} described in \cite{Ryan19}. This approach combines a decelerating spreading shell model \citep{vanEerten10} that includes a range of options for lateral and radial energy structure with the \textsc{emcee} (version 2.2.1) Python package for Markov-Chain Monte Carlo analysis \citep{Foreman-Mackey13}. For the jet model with a Gaussian distribution of lateral energy, the parameters are: fraction of post-shock internal energy in magnetic field $\varepsilon_B$, fraction of post-shock internal energy in the accelerated electron population $\varepsilon_e$, power-law slope of the electron population $-p$, homogeneous circumburst medium number density $n_0$, on-axis isotropic equivalent energy $E_0$, jet orientation $\theta_v$, jet core width $\theta_c$, and jet total width $\theta_w$.
We also perform fits that include an additional constant X-ray component, specified by a flux density $F_X$. This accounts for additional sources of emission, such as a long-lived engine or a separate source at close proximity on the sky.
We use the same prior on jet orientation as reported in earlier work \citep{Troja18}, drawn from \cite{Hubble170817} with a Hubble constant as determined by \cite{Planck18}.  The additional component $F_X$ is given a flat prior and bounded by $0 < F_X < 2 \times 10^{-4}$ $\mu$Jy. 

In order to explore the non-thermal emission from the sub-relativistic ejecta, we consider a quasi-spherical ``kilonova afterglow'' model.  While the bulk of the kilonova material coasts at a sub-relativistic velocity, it is expected a less massive tail of material outflows with substantially higher velocities \citep{Bauswein13,Hotokezaka13}.  The material is postulated to have an energy distribution which is a power-law in the four-velocity: $E_{>u}(u) = E_{\mathrm{tot}} (u/u_{\mathrm{min}})^{-k}$.  We use the same MCMC routines as with the structured jet analysis and the isotropic outflow model from \cite{Troja18}, reparameterized for a kilonova-like outflow.  
This model is specified by a power-law $k$ stratification of ejecta velocities, a total ejecta mass $M_{ej} = 2k/(k+2) u_{\mathrm{min}}^{-2} E_{\mathrm{tot}} c^{-2}$, a maximum ejecta four-velocity $u_{max}$, a minimum velocity $\beta_{\rm min}$, as well as the environmental and synchrotron parameters $n_0$, $p$, $\epse$, and $\epsB$. It is not a given that $\epse$, $\epsB$ and $p$ are identical for jet and kilonova component.

The structured jet fits used a parallel tempered ensemble MCMC sampler with 20 geometrically spaced temperatures between 1 and $10^6$. Each temperature rung was occupied by 100 walkers, and the chain was run for 20,000 iterations. The kilonova afterglow fits were run using a standard ensemble sampler with 300 walkers for 64,000 iterations. Further details of the method can be found in the references listed above.
Our models were compared to the X-ray, radio and optical afterglow light curves using the same data set described in \citep{Troja19}, and by adding the latest data from \citet{Mooley18b,Fong19,Hajela19} and this work.

To compare different models we utilize the Widely Applicable Information Criterion (WAIC; \citealt{Watanabe10}).  The WAIC is an estimate of the ``expected log predictive density'' (\emph{elpd}): a score measuring the likelihood new data will be well described by the current model  \citep{Gelman13}.  The \emph{elpd} measures the predictive power of a fit, it rewards a tight match to the data while penalizing over fitting and extraneous parameters.  The WAIC is proven to be asymptotically equal to the \emph{elpd} for a wide range of models and is straightforward to compute from MCMC posterior samples, whereas the \emph{elpd} itself can only be computed if the true model is known.  We use the $p_{\mathrm{WAIC} 2}$ estimator for the effective number of parameters \citep{Gelman13}.

Following \citet{Vehtari17} we compute the WAIC score for each model at every data point.  The total WAIC score WAIC$_{elpd}$ for a model is the sum of the scores for each data point.  Each model score and score difference $\Delta$WAIC$_{elpd}$ have a standard error computed from the variance over the contributions from each data point.  This standard error is likely optimistic but within a factor of 2 of the true value \citep{Bengio04}.  In a two-way comparison, a model is favoured if its $\Delta$WAIC is several times larger than its standard error.

\newpage

\section{Results}

Two and a half years after the merger, 
{\it Chandra} continues to detect X-ray emission at the location of GW170817. 
A comparably long-lived X-ray emission is rare in GRBs, and was reported only for long duration bursts, such as GRB~130427A \citep{DePasquale16} and GRB~980425 \citep{Kouveliotou04}. 
For a spectral index $\beta$=0.585, the extrapolation of the observed X-ray emission
corresponds to $F606W$\,$\approx$29.7$\pm$0.3 AB mag in the optical and $\approx$5$\pm$2~$\mu$Jy at 3~GHz. For comparison, at the GW location \textit{HST}/WFC3 can reach a 5\,$\sigma$ point-source sensitivity of $F606W$\,$\approx$28 AB mag in four orbits \citep{Lamb19},  whereas a 6~hr long VLA observation can reach a 5\,$\sigma$ sensitivity of $\approx$10-15\,$\mu$Jy in S-band\footnote{https://obs.vla.nrao.edu/ect/}. 
X-ray observations therefore remain the most
powerful probe into the faintest stage of the GW counterpart.

In the latest epochs, the measured X-ray flux is higher than model predictions based 
on the earlier dataset \citep{Troja19}, suggesting a shallower temporal decay. 
Contamination from an unrelated X-ray source seems unlikely. 
The probability of a background AGN of comparable flux is about $ 10^{-4}$\,arcsec$^{-2}$ \citep{Georgakakis08}. The density of luminous X-ray sources within the galaxy is also relatively small,
as can be directly seen from Fig.~\ref{fig:image}. 
The population of X-ray binaries in elliptical galaxies is in part associated to globular clusters, 
however deep {\it HST} observations find no globular cluster at the transient position \citep{Troja17,Lamb19}.
The density of field X-ray binaries depends on the specific star formation rate (sSFR).  Present systematic studies cover the range of  log (sSFR)$>-12.1$ \citep{Lehmer19}, while NGC~4993 has a much lower value, log(sSFR$<-13$) \citep{Im17}. 
Assuming that the relationship established at higher vales of sSFR holds,  $\lesssim\,10$ X-ray binaries with $L_X\gtrsim$3$\times$10$^{38}$\,erg\,s$^{-1}$ are expected in NGC~4993. Taking into account the distribution of X-ray sources as a function of their radial offset \citep{Mineo14}, we derive a chance alignment of $\approx10^{-3}$ arcsec$^{-2}$ at the position of GW170817.

Any significant departure from the jet model is likely inherent to the source, and could be caused by several factors, which we discuss below.

\begin{figure}
\includegraphics[width=1.0\columnwidth]{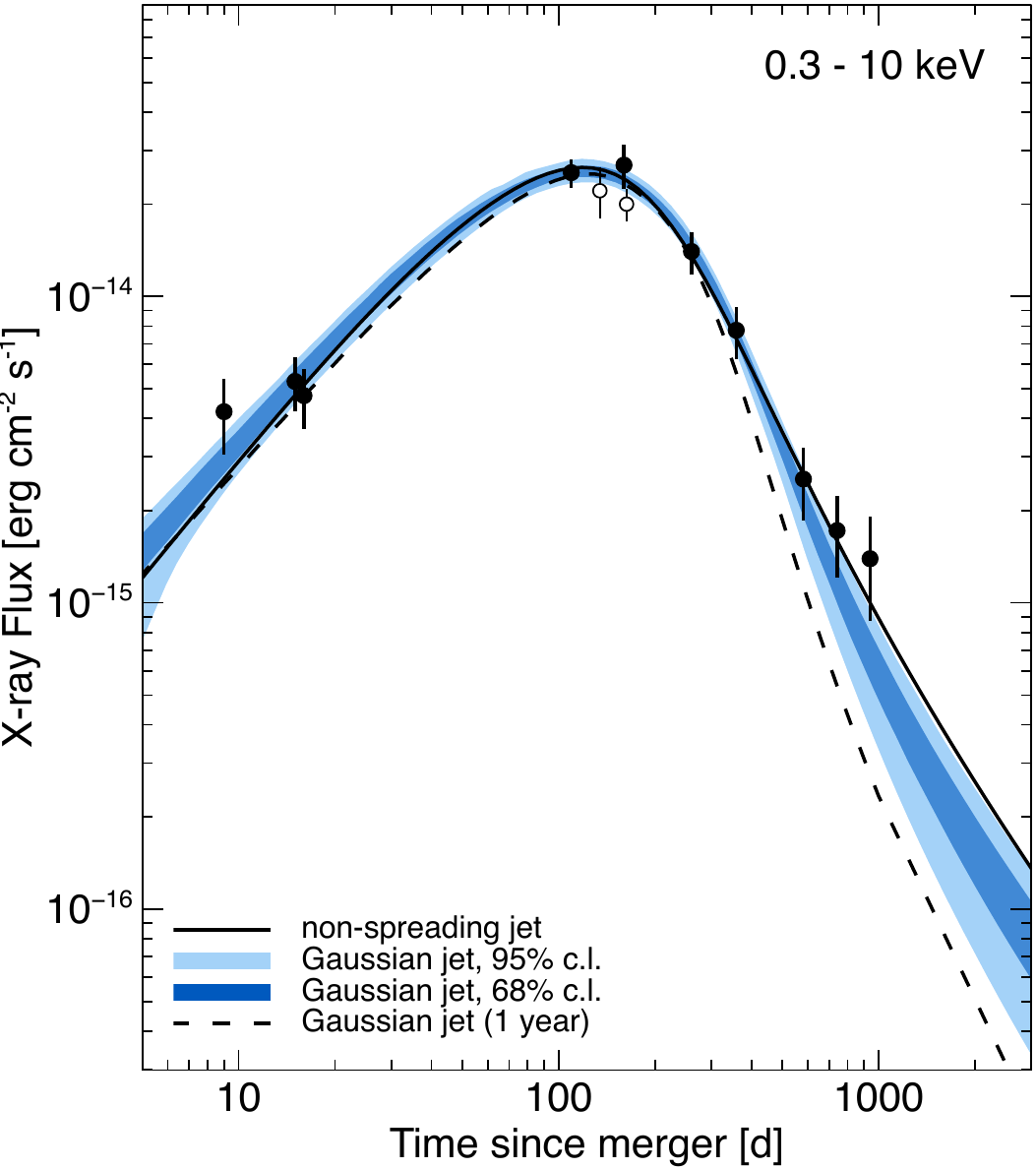}
\vspace{-0.3cm}
\caption{X-ray afterglow light curve of GW~170817, including 
\textit{Chandra} (filled circles) and \textit{XMM-Newton} (open circles) measurements.
The dashed line shows the best fit results from earlier work \citep{Ryan19}, based on the first year of data \citep{Troja19}.
The dark (light) blue range shows the 68\% (95\%) uncertainty region of the updated fit, including the entire dataset.
The solid line shows the best fit non-spreading jet model.
}
 \label{fig:jet_only_fits}
\end{figure}

\subsection{Jet}

Figure \ref{fig:jet_only_fits} compares the X-ray dataset to the range of jet model light curve predictions. The fit results are summarized in Table \ref{tab:fit_results_jet}.
Our previous best fit \citep{Ryan19}, based on the full first year data set, is shown by the dashed curve. The discrepancy between the new data and these earlier predictions is approximately $2 \sigma$, with the previous fit notably under-predicting the new observations.
A refit of the full updated data set is also shown in Fig. \ref{fig:jet_only_fits}, the solid bands denoting the distribution of X-ray flux estimated by the model. Even though the new fit result curve intersects the new observations within their error bars, it is nevertheless of interest that the model still systematically under predicts the late time observations.  Updated posterior parameter constraints are shown in Table \ref{tab:fit_results_jet}. The new constraints are consistent within the uncertainties with those from the first year data, although both the viewing angle $\thetaobs$ and circumburst density $n_0$ center on higher values than before.

Both these increases can be understood on simple grounds.  The early rise of the jet fixes the ratio $\thetaobs/\theta_c$ but leaves their absolute values relatively unconstrained \citep{Ryan19, np20}.  As the jet is slowly approaching the Sedov regime, the brighter than expected late X-ray emission requires a wide jet to contribute more flux.  Indeed, Table \ref{tab:fit_results_jet} shows our fit value for the opening angle $\theta_c$ increased from 0.07~rad to 0.09~rad when the new observations were included.  Since the early afterglow fixes $\thetaobs / \theta_c$, the required viewing angle increases as well.  The circumburst density is increased to keep the jet break at 160~d, compensating for the increased viewing angle which would otherwise push the jet break to a later time \citep{Ryan19}.

In Table~\ref{tab:fit_results_jet}, we compare the results of our modelling to additional observing constraints, which were not input into the fit. \cite{Ghirlanda19} constrained the size of the radio centroid at $T_0+207$~d to $\delta < 2.5$ mas at 90\% confidence.  All our models are safely within this limit.  
A more stringent constrains comes from the apparent velocity $\beta_{\textrm{app}}$ of the center-of-brightness on the sky. A value of $\beta_{\textrm{app}} = 4.1 \pm 0.5$ has been obtained from Very Large Baseline Interferometry (VLBI) by \cite{Mooley18c}, measured between 75~d and 230~d after the burst.  The model fit to the first year of data estimates $\beta_{\textrm{app}} = 3.5^{+1.2}_{-0.8}$, consistent with the observed value.  However, the updated fit significantly under predicts the observed centroid movement, estimating only $\beta_{\textrm{app}} = 2.2^{+0.5}_{-0.4}$.  This is largely due to the increased viewing angle, to which the superluminal apparent velocity is a sensitive function.

\begingroup
\renewcommand{\arraystretch}{1.5}
\begin{table*}
         \caption{Fit results for the jet models. Col.~1 reports the parameters name and units. 
         Col.~2: a Gaussian structured, spreading jet fit to the first 360 days of observations.
Col.~3: identical jet model fit to all 940 days of data.
Col.~4: a Gaussian jet with spreading artificially stopped. This model is not physical, but serves to bracket the diversity of possible behaviours of spreading jets.
Col.~5: a spreading Gaussian jet with an additional constant X-ray flux.
         }
        \begin{tabular}{lrrrr}
        \hline
        \multirow{3}{*}{Parameter} & 
        \multicolumn{1}{c}{360 d} & \multicolumn{3}{c}{940 d}\\
         \cmidrule(lr){2-2} \cmidrule(l){3-5}
         & Spreading Jet & Spreading Jet & Non-spreading    & Spreading Jet\\
        &                &               & Jet      & Plus Constant \\
         \hline
  $\theta_v$ (rad)       
     & $0.40^{+0.11}_{-0.11}$    
     & $0.54^{+0.09}_{-0.10}$ 
     & $0.31^{+0.08}_{-0.08}$ 
     & $0.44^{+0.10}_{-0.11}$ \\
  $\log_{10} E_0$ (erg)       
  & $52.9^{+1.0}_{-0.7}$   
  & $53.0^{+0.90}_{-0.90}$   
  & $53.2^{+1.0}_{-0.8}$    
  & $53.2^{+1.0}_{-1.0}$ \\
  $\theta_c$ (rad)      
  & $0.07^{+0.02}_{-0.02}$    
  & $0.088^{+0.014}_{-0.015}$ 
  & $0.047^{+0.011}_{-0.011}$ 
  & $0.071^{+0.017}_{-0.018}$ \\
  $\theta_w$ (rad)       
  & $0.47^{+0.30}_{-0.19}$    
  & $0.6^{+0.3}_{-0.3}$    
  & $0.34^{+0.18}_{-0.14}$    
  & $0.5^{+0.3}_{-0.2}$ \\
  $\log_{10} n_0$  (cm$^{-3}$) 
  & $-2.7^{+1.0}_{-1.0}$   
  & $-1.7^{+0.9}_{-1.0}$   
  & $-2.7^{+1.1}_{-1.1}$      
  & $-2.3^{+1.1}_{-1.1}$ \\
  $p$                   
  & $2.170^{+0.010}_{-0.010}$    
  & $2.139^{+0.010}_{-0.010}$    
  & $2.160^{+0.009}_{-0.017}$   
  & $2.146^{+0.012}_{-0.011}$ \\
  $\log_{10}\epse$            
  & $-1.4^{+0.7}_{-1.1}$   
  & $-2.0^{+0.8}_{-0.8}$   
  & $-1.9^{+0.8}_{-1.1}$    
  & $-2.1^{+0.9}_{-1.0}$ \\
  $\log_{10}\epsB$            
  & $-4.0^{+1.1}_{-0.7}$   
  & $-3.7^{+0.9}_{-0.9}$   
  & $-3.8^{+1.1}_{-0.9}$    
  &  $-3.4^{+1.0}_{-1.0}$ \\
      \hline
  $E_{\mathrm{tot}}$  (erg) 
  & $50.6^{+0.9}_{-0.7}$ 
  & $50.9^{+0.9}_{-0.9}$ 
  & $50.5^{+1.0}_{-0.8}$ 
  & $50.8^{+0.9}_{-0.8}$ \\
  $\beta_{\mathrm{app}}$ ($c$)    
  & $3.5^{+1.2}_{-0.8}$ 
  & $2.2^{+0.5}_{-0.4}$ 
  & $4.3^{+1.4}_{-0.9}$ 
  & $2.7^{+1.0}_{-0.6}$ \\
  $\delta_{\mathrm{rms}}$ (mas)  
  & $0.60^{+0.3}_{-0.14}$ 
  & $0.61^{+0.12}_{-0.09}$ 
  & $0.48^{+0.16}_{-0.10}$ 
  & $0.75^{+0.3}_{-0.14}$ \\
  \hline
  $\tilde{\chi}^2$ (dof)        
  & 1.51 (94)                        
  & 1.20  (94)                    
  & 1.29  (94)                     
  & 1.18 (93) \\
  WAIC$_{\mathrm{elpd}}$          
  & --                        
  & $694.5$          
  & $690.4$             
  & $695.7$ \\
  $\Delta$ WAIC$_{\mathrm{elpd}}$  
  & --               
  & 0.0                       
  & $-4.1 \pm 3.1$              
  & $1.2 \pm 1.4$ \\ 
  \hline
    \end{tabular}\\
    \justify
\textbf{Notes -} Marginalized posterior values for each fit parameter, the median and 68\% confidence interval, from the MCMC runs are given in columns 2 - 5, rows 1-8.  
Rows 9-11 give the marginalized posterior values for the total energy $E_{\mathrm{tot}}$, apparent velocity $\beta_{\mathrm{app}}$ measured between the VLBI observations \citep{Mooley18c}, and rms width of the centroid during the EVN observations \citet{Ghirlanda19} respectively, also with median and 68\% confidence interval. 
The last three rows give the reduced $\tilde{\chi}^2$ value of the maximum-posterior estimate (and degrees of freedom for each fit), the WAIC estimate of the expected log predictive density (elpd), and difference between the WAIC values and the spreading Gaussian Jet fit with standard error.
A higher elpd indicates a model better able to predict the data.
   \label{tab:fit_results_jet}
\end{table*}
\endgroup

In our Gaussian jet model the observed motion of the radio afterglow centroid, which requires smaller viewing angles, appears therefore in slight tension with the late X-ray flux, which instead favors larger viewing angles.  The tension could be alleviated if the afterglow light curve were able to flatten faster than our current modelling allows. Such an effect could originate from the dynamics of the GRB jet, changes to the emitted synchrotron spectrum, or possibly an additional emission component.

Because the spreading of GRB jets occurs during an intermediate dynamical regime between ultra-narrow highly relativistic flow and broad non-relativistic flow, the evolution of the jet during the spreading stage is more sensitive to the details of outflow geometry than either asymptotic limit of behaviour would suggest. This affects both multi-dimensional hydrodynamical simulations of jets and semi-analytical models. Our model is based on a semi-analytical model for jet spreading \citep{Ryan19, vanEerten10}, and shares this sensitivity. For that reason, we also test the extreme assumption of no spreading at all.  Such a jet is non-physical, but serves to bracket the range of jet model light curve predictions.

We ran a fit to the full dataset with a non-spreading Gaussian jet. The best fit (maximum posterior) light curve is shown in Figure~\ref{fig:jet_only_fits} (solid line) and the summary of fit results are presented in Table~\ref{tab:fit_results_jet}.  The non-spreading jet has a slower decay after the jet break and is more easily able to accommodate the late data points while requiring an earlier and broader peak. Changing the model assumption about jet spreading mostly affects our inferred values for the angles and circumburst density (see Table~\ref{tab:fit_results_jet}).  These end up smaller, consistent with the previous estimates derived from the dataset at 360~d but outside the uncertainties from the fit to the full dataset.  The apparent velocity increases to $\beta_{\textrm{app}}=4.3^{+1.4}_{-0.9}$ due to the smaller viewing angle, and is consistent with the observed value.  
Although this model does not describe a realistic jet configuration, this fit serves to demonstrate that the interpretation of afterglow data at these late times is highly sensitive to the dynamics of jet spreading.

Both for jets with and without lateral spreading, the full transition to the non-relativistic regime takes 
$t_{\rm NR}$\,$\approx$\,$10^4$ days to complete and will not impact the light curve at  the current time scale of observations for a reasonable range of model parameter values. The same holds for the appearance of the counter jet, which our models project to temporarily lead to a near-flat light curve between 3000-5000 days after the burst (at around 10$^{-16}$\,erg\,cm$^{-2}$\,s$^{-1}$ at X-ray frequencies and around 0.2\,$\mu$Jy at 3 GHz).

Rather than the divergence between model and data being due to limitations of the model, the jet dynamics might also genuinely change under changing external conditions, specifically a change in circumburst density. Analytical modeling for a homogeneous environment show that the flux below the cooling break scales proportional to circumburst number density $n$ according to $n^{1/2}$ and $n^{0.4}$ ($p=2.2$) in the relativistic and non-relativistic limit respectively (see e.g. \citealt{Leventis12}). In other words, it would merely take a factor four increase in density at distances beyond about a parsec from the merger site (the approximate distance traveled by the jet when observed at its light curve peak around 160 days) in order for the light curve baseline to drift towards a factor two increase, consistent with the latest observations.

A change in the light curve slope 
can also occur if the  synchrotron cooling break frequency enters or exits the X-ray band. However, our structured jet modeling shows that both radio and X-ray light curves remain in the same spectral regime between injection break $\nu_m$ and cooling break $\nu_c$ throughout our observations, and that $\nu_c$ shifts upwards again after a closest approach to the X-ray band during the light curve peak around 160 days.
This is exactly the same evolutionary pattern for $\nu_c$ as predicted across jet breaks from ultra-high resolution numerical hydrodynamics simulations (starting from top-hat initial conditions, see Fig~4 of \citealt{vanEertenMacFadyen13})
and matches the evolution of the hardness ratio (Figure~\ref{fig:hr}), 
although, given the large error bars, it is not possible to draw too strong a conclusion about this similarity.
We therefore find it unlikely that the cooling frequency $\nu_c$ affects the latest X-ray observations. 

Finally, it could be the case that the synchrotron parameters themselves evolve over time. For example, the value of $p$ evolving closer to 2, as expected for non-relativistic shock speeds \citep{Blandford78, Bell78},  would indeed lead to a harder spectrum (from 0.585 for $p=2.17$ to 0.5 for $p = 2$) and shallower temporal slope (since $\alpha$ in $t^{-\alpha}$ equals $p$ for a fast spreading jet, $3p/4$ for a non-spreading jet and $(15p - 21)/10$ in the non-relativistic limit, see \citealt{ZhangMeszaros04, PK04,  Frail00}, respectively). 
However, this would also affect the overall flux normalization, which contains an $\varepsilon_e (p-2)$ term. Although the impacts of these effects on the light curve will be mitigated by the spread in emission arrival times from the blast wave, it would still require $\varepsilon_e$ to co-evolve such that a substantial shift in baseline flux level is to be avoided.
An updated broadband measurement of the slope of $p$ could directly answer the question whether $p$ is indeed evolving, but for now we conclude that the tentative flattening of the light curve has not been established as a generic prediction of such a scenario.

\begin{figure}
    \centering
    \includegraphics[width=0.98\columnwidth]{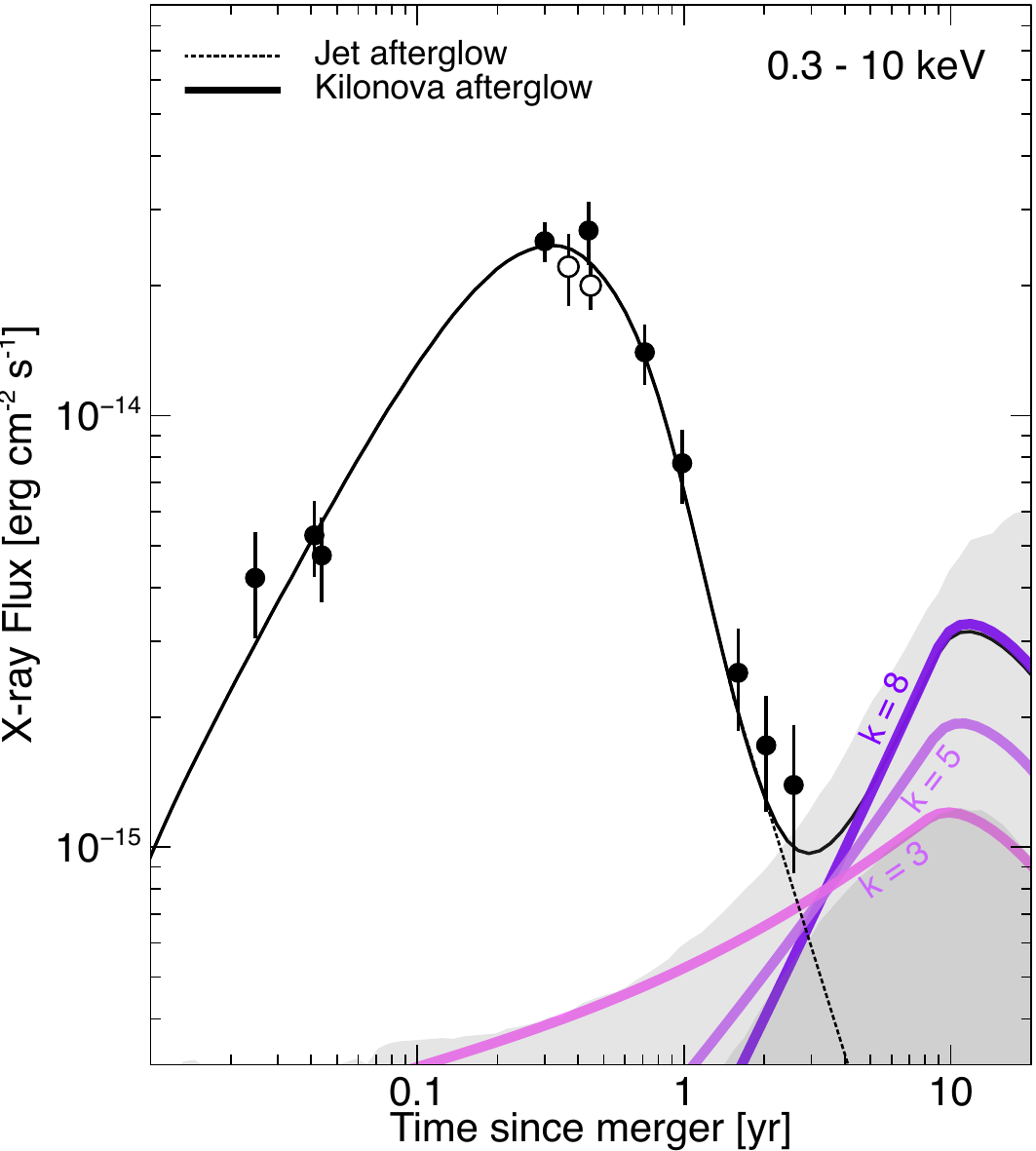}
    \vspace{-0.0cm}
    \caption{X-ray afterglow described with a Jet+Kilonova afterglow model (thin solid line), derived from broadband fitting.  The shaded gray area show the range of X-ray fluxes estimated by the model (light gray: 95\% c.l., dark gray: 68\% c.l.).
    The dotted line shows the contribution of the jet component, 
    whereas the thick solid lines show the evolution of the kilonova afterglow for different velocity indices $k$. 
    The three kilonova models were generated for the same set of input parameters ($M_{\rm ej}$ = 0.025\,$M_{\odot}$, 
    $\beta_{\rm min}$ = 0.3$c$,
    $p$ = 2.01, 
    $n$ = 8$\times$10$^{-3}$\,cm$^{-3}$,
    $\epsB$ = 6$\times$10$^{-5}$) and three different pairs
    of values ($k$=8,$\epse$=0.17; top), ($k$=5,$\epse$=0.089; middle), 
    and ($k$=3,$\epse$=0.045; bottom). 
    }
    \label{fig:KNlc}
    \vspace{0.4cm}
    \includegraphics[width=0.98\columnwidth]{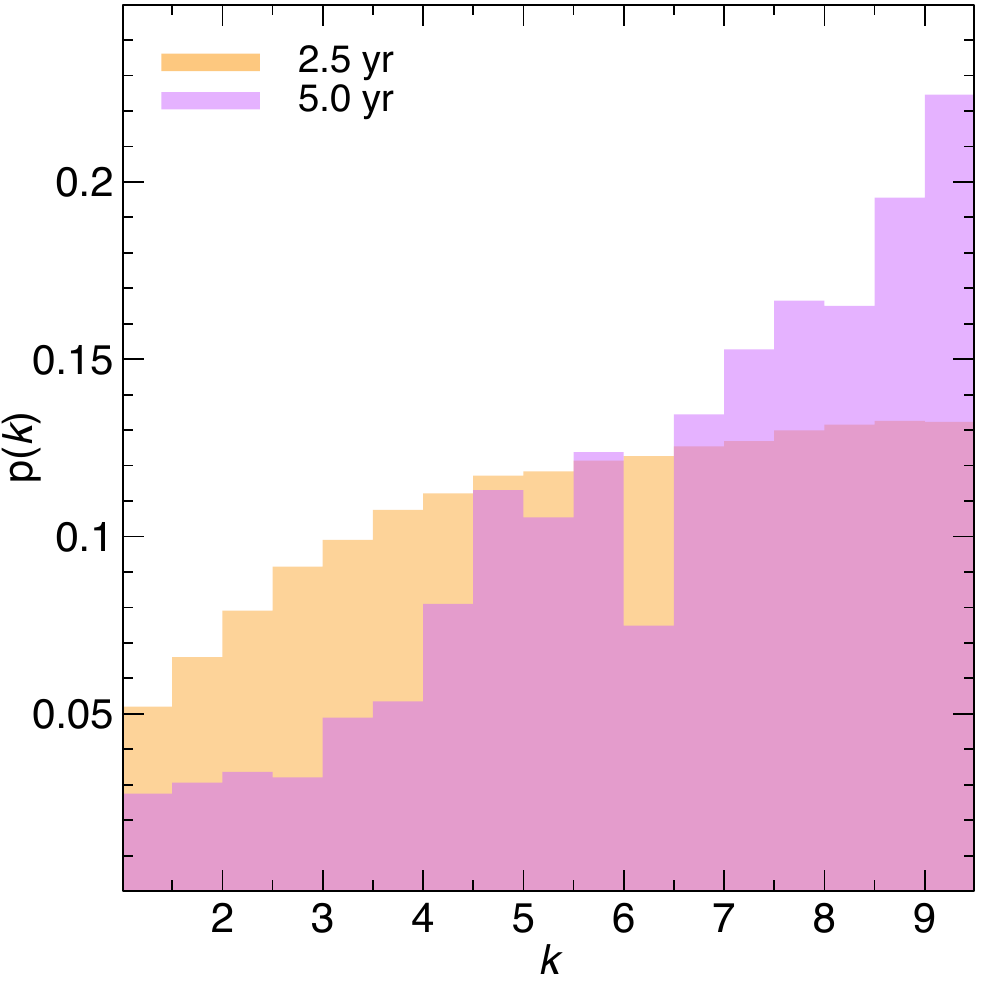}
        \vspace{-0.1cm}
    \caption{Posterior distribution on the ejecta velocity index $k$, assuming the kilonova afterglow  contributes to the observed X-ray flux at 2.5~yrs (orange).  Radio upper limits were also included in the fit. In purple, the posterior distribution on $k$ if the X-ray flux remains above the current level up to 5 years after the merger.   
    }
    \label{fig:KNkdist}
\end{figure}

\subsubsection{Limits On An Additional Component}\label{sec:add}

Table \ref{tab:fit_results_jet} presents the results of fitting an additive constant X-ray flux to the spreading, Gaussian structured jet afterglow.  In such a scenario the viewing angle $\thetaobs$ and circumburst density $n_0$ are somewhat reduced compared to their values from the jet alone, and consistent with the values
derived from the 1 year dataset. 
The additional flux density at 5 keV is constrained to $F_X = (2.8 \pm 1.2)\times 10^{-5} \mu$Jy, corresponding to (8 $\pm$ 3) $\times$ 10$^{-15}$ erg\,cm$^{-2}$\,s$^{-1}$, about half the observed flux at $T_0+939$ d.   The smaller viewing angle causes a larger apparent velocity, $\beta_{\mathrm{app}} = 2.7^{+1.0}_{-0.6}$, consistent with the observations of \citet{Mooley18c}.
The improvement in WAIC score between the jet plus constant and the standard jet is marginal  ($1.2 \pm 1.4$), and does not warrant the addition of another parameter in the model.

\subsection{Kilonova Afterglow}

We use the latest X-ray and radio observations  
to constrain the range of valid kilonova afterglow models\footnote{For simplicity, we only discuss kilonova models that do not invoke additional energy injection from a long-lasting central engine \citep[e.g.][]{gao13}}.
In lieu of running a combined fit with both structured jet and kilonova afterglows, we use the structured jet plus constant fit (Sect.~\ref{sec:add}) as a measure of the possible contribution of the kilonova afterglow to the current epoch. 
We run a simple MCMC fit with the kilonova afterglow model to the X-ray flux $F_X\approx8\times10^{-15}$\,erg\,cm$^{-2}$\,s$^{-1}$, as well as the latest radio upper limits.  There are no other constraints apart from priors and the requirement the light curve be currently rising.

We focus our study on the emission arising from the fastest ejecta, often referred as the ``blue'' kilonova component, as it is expected to peak ealier and initially be brighter \citep{Alexander17, Kathirgamaraju19}.
Our prior on $M_{ej}$ is a normal distribution with mean $2.25 \times 10^{-2} M_{\odot}$ and width $0.75 \times 10^{-2} M_{\odot}$,
 as derived from the modeling of AT2017gfo \citep[e.g.][]{Arcavi2017,Evans2017,Nicholl2017, Kasen17,Pian2017,Tanvir2017,Troja17}. 
Our prior on the minimum outflow velocity $\beta_{\mathrm{min}}$ is a normal distribution with mean $0.3$ and width $0.05$, as lower values would lead to delayed and dimmer peaks below our detection limits \citep{Kathirgamaraju19}. 
The velocity distribution index $k$ was given a uniform prior between $1$ and $10$. 
The circumburst density $n_0$ was given a log-uniform prior between $10^{-3}$ and $10^{-1}$ cm$^{-3}$ in agreement with the constraints from the jet model.  
The electron spectral index $p$ was given a uniform prior between 2 and 3, while $\epse$ and $\epsB$ were given log-uniform priors between $10^{-5}$ and 1. We note these parameters are under no obligation to take identical values in both the structured jet and kilonova afterglow.  

We find the current data set admits a broad range of kilonova models (Figure~\ref{fig:KNlc}) and is insufficient to provide strong constraints on any of the parameters, including the velocity distribution index $k$.  Figure \ref{fig:KNkdist} shows the posterior probability distribution on $k$.  Essentially any value is consistent with current observations.  Preliminary constraints (disfavoring $k$\,$<$6) were derived by \citet{Hajela19}, our exploration of the parameters space finds instead a broader range of possible solutions. 
This result is consistent with the analysis presented in 
\citet{Hajela19}, in particular their Fig.~5 showing a wide range 
of allowed values, but does not support the conclusion $k$\,$\geq$6. 
Higher values of $k$ result in fainter initial emission and a steep rise to the peak flux.  Lower values of $k$ are instead brighter at earlier times with a slow rise to the final flux.  These are easily brought in agreement with the current observations with a slight reduction of $\epse$ and $\epsB$ (Figure~\ref{fig:KNlc}).
Continued monitoring of this target would therefore be critical to determine the rising slope of the kilonova afterglow component, and constrain the ejecta velocity profile.

Unfortunately, due to the large number of parameters and uncertainty in the physical properties of the kilonova blast wave, it is difficult to make robust conclusions about its afterglow emission at this time.  
Ultimately, the large uncertainty in the synchrotron parameters $\epse$ and $\epsB$ dominate the analysis, and will only be overcome with successful observations.
As shown in Figure~\ref{fig:predictions}, the same observing settings thus far adopted to monitor GW170817 probe the top 30\% of the estimated flux distribution and
could detect the kilonova afterglow under favorable conditions.

\subsection{Energy injection from a pulsar}

Another possibility of flattening the light curve is to invoke energy injection of a long-lived NS. This possibility was suggested by \cite{Piro19} to account for the X-ray variability around 160 days and to interpret some of the features in the kilonova AT2017gfo associated with GW170817 \citep{Yu18,Li18,Wollaeger19}. A long-lived NS central engine is allowed by the EM and GW observational data, as long as the surface dipole field strength is not very strong \citep{Ai18} and the NS equation of state is stiff enough \citep{Ai19}. For such a NS, the spindown time scale can be of the order of years, so that significant energy injection is still possible at the time of our observations. Indeed, \cite{Piro19} predicted the flattening of the lightcurve based on their model parameters to interpret the X-ray variability. 

\begin{figure}
\includegraphics[width=0.98\columnwidth]{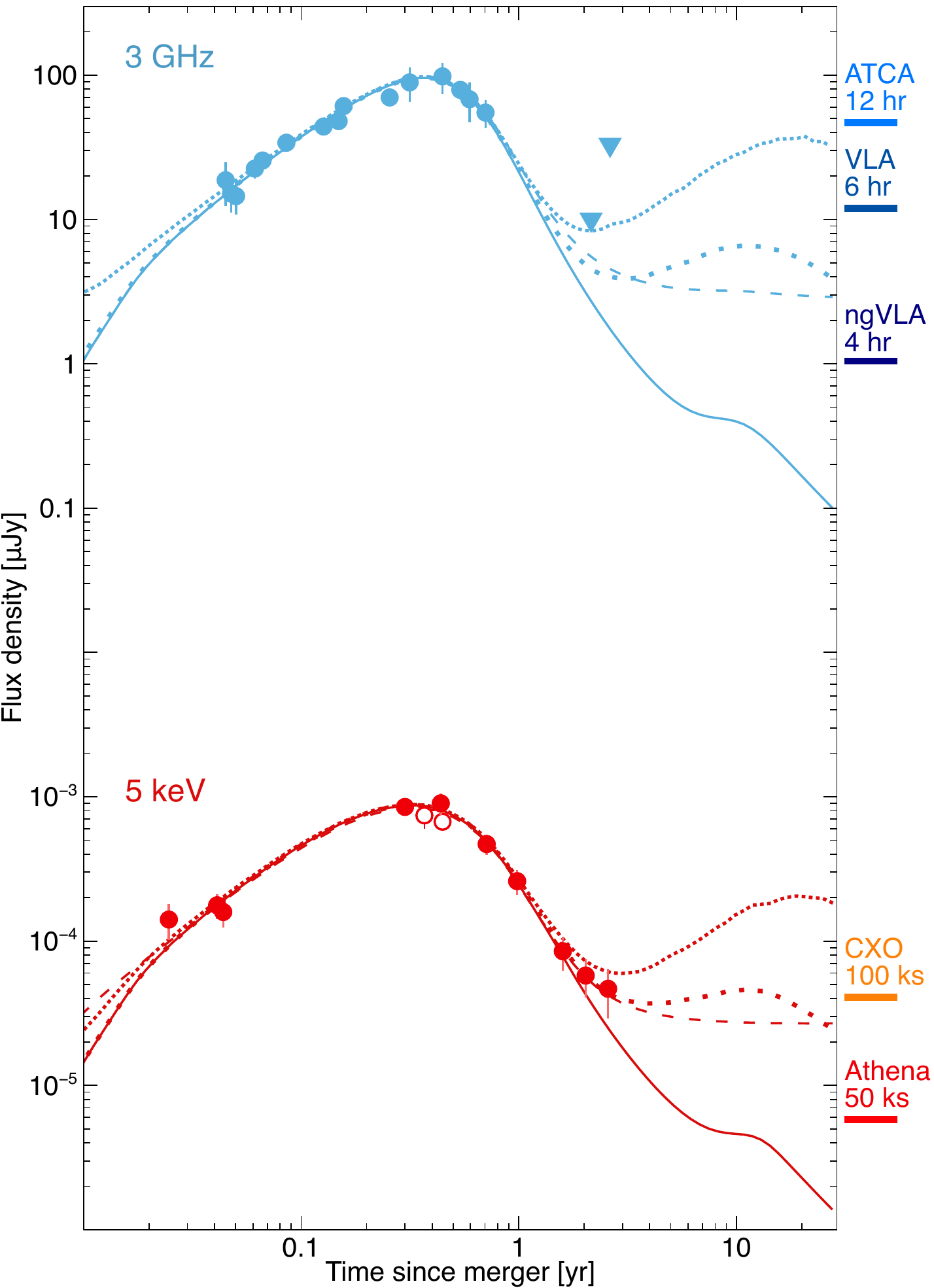}
\caption{X-ray and radio light curves of GW170817, showing the possible future evolution of the emission components: 
the relativistic jet (solid line), the kilonova afterglow (dotted lines) , and the remnant neutron star (dashed line). Emission from the counter-jet causes a flattening of the jet lightcurve at $\approx$10 years after the merger. 
For the kilonova afterglow, we report the two upper bounds (95\% and 68\% confidence intervals) on the estimated flux distribution (cf. Figure~\ref{fig:KNlc}). 
On the right we report the 5\,$\sigma$ sensitivity of typical observing settings
for ATCA, VLA, and \textit{Chandra} (CXO), as well as 
for the next generation ngVLA \citep{CorsiWP} and the Athena X-ray observatory. }
 \label{fig:predictions}
\end{figure}

We consider a general energy injection law from the central engine, 
$L(t) \propto t^{-q}$,
where $q<1$ is needed to give a noticeable change of blastwave dynamics \citep{Zhang01}. We consider two possibilities. The first is that the spindown luminosity is injected into the blastwave as a Poynting flux. For GW170817/GRB 170817A, the current epoch is already in the post-jet-break phase since the light curve is already in the rapid decay regime. Let us assume that the blastwave is still in the relativistic regime and that sideways expansion is not important, one can derive an analytical model for decay slopes. For a constant density medium (which is relevant for NS-NS mergers), one has \citep{Zhang18}\footnote{The relevant parameters are for the jet core and an on-axis observer. For a structured jet with a large viewing angle like the case of GRB 170817A, these scalings are relevant after the jet core enters the line of sight, i.e. during the rapid decay phase.} $\Gamma \propto t^{-(2+q)/8}$, $r \propto t^{(2-q)/4}$, $\nu_m \propto t^{-(2+q)/2}$, $\nu_c \propto t^{(q-2)/2}$. The peak flux density can be estimated as $F_{\rm \nu,max} \propto r^3 B' \Gamma [\theta_j^2 / (1/\Gamma)^2] \propto r^3 \Gamma^4 \propto t^{(2-5q)/4}$. For $\nu_m < \nu < \nu_c$ which is relevant for X-rays at such a late epoch, the flux density evolution should satisfy
\begin{equation}
F_\nu \propto t^{\frac{2-5q}{4}-\frac{(p-1)(2+q)}{4}}.
\end{equation}
This expression is consistent with the pre-jet-break energy injection theory \citep{Zhang06} if the edge effect correction factor $[\theta_j^2 / (1/\Gamma)^2]$ is removed. For $q=0$ relevant to pulsar injection in the pre-spindown phase, this gives $F_\nu \propto t^{(2-p)/2}$, which is nearly flat (for our best fit $p = 2.17$, this gives $t^{-0.085}$). This is consistent with the numerical result presented in Figure~\ref{fig:predictions}. For this first scenario, energy injection should be achromatic. The same flattening feature should appear in the radio band as well.

The second scenario invokes an internal dissipation of the pulsar wind, which has been manifested by the so-called ``internal plateaus'' as observed in both long \citep{Troja07,Lvzhang14} and short \citep{rowlinson10,lv15} GRBs. The temporal profile should directly follow $\propto t^{-q}$, which is also flat for $q=0$. The light curve should be chromatic, as seen in GRB afterglows \citep{Troja07,rowlinson10,Lvzhang14,lv15}, and the radio band may not show a simultaneous flattening as the X-ray band. Since all the other flattening mechanisms (discussed earlier in Sect.3.1 and 3.2) also predict achromatic behaviors, a detection of chromatic behavior between X-ray and radio will provide a definite clue about a long-lived central engine. 

If the flattening is indeed caused by energy injection of a long-lived pulsar, the spindown time scale should be at least this long, i.e. \citep{dai98,Zhang01}
\begin{equation}
    T_{\rm sd} \sim (2\times 10^7 \ {\rm s}) \ B_{p,13}^{-2} P_{0,-3}^2 > 1,000 \ {\rm d},
\end{equation}
where $B_p = 10^{13} \, {\rm G} \ B_{p,13}$ is the surface polar magnetic field strength, and $P_0 = 1 \, {\rm ms} \ P_{0,-3}$ is the initial spin period of the pulsar. This condition is readily satisfied if $B_p$ is below a few times of $10^{12}$~G, which is consistent with the constraints from other observations from this event \citep{Ai18,Ai19,Piro19}. Within the energy injection model, lightcurve flattening appears when the injected energy exceeds the original energy in the blastwave, and the ceases when the total available spin energy is injected. According to our structured jet modeling, the total kinetic energy in the jet is $\sim 10^{50}-10^{52}$ erg with medium value $5\times 10^{50}$ erg. This is smaller than the typical available spin energy of a new-born millisecond pulsar from an NS-NS merger (typically a few $10^{52}$ erg, but could be smaller due to possible a secular gravitational wave loss, \citealt{fan13,gao16}). As a result, such an energy injection is expected if the merger product is indeed a long-lived neutron star. The injection energy may be up to a factor of a few to a few hundreds of the existing energy in the jet, so that the injection episode may last for years according to this model.

\section{Conclusions}

Whereas optical and radio emission from GW170817 have now faded below detection threshold, its X-ray counterpart continues to be visible at 2.5 years after the NS merger. Earlier predictions of the structured jet model systematically underestimate the latest {\it Chandra} detections. A Gaussian structured jet can still reproduce the afterglow temporal evolution by increasing the viewing angle to $\approx$30$^{\circ}$, although this updated model underpredicts the centroid motion, as constrained by high-resolution radio imaging. Alternatively, the slow X-ray decline could indicate a genuine new feature 
of the afterglow, originating from the dynamics of the GRB jet, changes to the emitted synchrotron spectrum, or possibly an additional emission component. The latter contribution is constrained by our modeling to $F_X$\,$\approx$8$\times$10$^{-15}$\,erg\,cm$^{-2}$\,s$^{-1}$,
corresponding to an X-ray luminosity  
$L_X$\,$\approx$1.5$\times$10$^{38}$\,erg~s$^{-1}$ (0.3-10~keV).

Continued energy injection by a long-lived central engine would cause a persistent flattening of the X-ray lightcurve. Depending on the origin of this emission (internal or external), the same flattening could be observed in the radio band.  The observed behavior could mark instead the onset of a non-thermal ``kilonova afterglow", produced by the interaction of the sub-relativistic merger ejecta with the surrounding medium.  
We find that the current dataset is not sufficient to meaningfully constrain any of the parameters, including the velocity distribution index $k$. Our results do not support earlier predictions 
of $k$\,$\geq$\,6 and find a wide range of allowed values.
Future multi-band observations of this component would be essential to determine the velocity profile of the sub-relativistic ejecta, thus complementing earlier kilonova studies, based on the thermal optical/nIR emission.

\section*{Acknowledgements}

The authors wish to thank the ATCA staff for the support in carrying out the observations during the current health emergency.
ET thanks A. Hornschemeier and A. Basu-Zych for their
helpful feedback, and is grateful to B. A. Vekstein 
for her fundamental cooperation during the writing of this manuscript. 
This work was partially supported by the National Aeronautics and Space Administration through Chandra Award Number G0920071A issued by the Chandra X-ray Center, which is operated by the Smithsonian Astrophysical Observatory for and on behalf of the National Aeronautics Space Administration under contract NAS8-03060."
GR acknowledges the support from the University of Maryland through the Joint Space Science Institute Prize Postdoctoral Fellowship.  Analysis was performed on the YORP cluster administered by the Center for Theory and Computation, part of the Department of Astronomy at the University of Maryland. 
LP acknowledges partial support by  the
European Union Horizon 2020 Programme under the AHEAD2020 project (grant
agreement number 871158).



\bibliographystyle{mnras}


\label{lastpage}
\end{document}